# Anomalous Thermoelectric power of over-doped $Bi_2Sr_2CaCu_2O_8$ superconductor

V.P.S. Awana[1,*], Jagdish Kumar Bains[1], G.S. Okram [2], Ajay Soni[2] and H. Kishan[1]

[1]National Physical Laboratory, Dr. K.S. Krishnan Marg, New Delhi – 110012, India.
[2]UGC-DAE Consortium for Scientific Research, Indore 452001, India

Temperature dependence of thermoelectric power S(T) of three differently processed $Bi_2Sr_2CaCu_2O_8$ (Bi2212) samples, viz. as-processed melt quenched (Bi2212-MQ), $600^0$C $N_2$-annealed (Bi2212-$N_2$) and $600^0$C $O_2$-annealed (Bi2212-$O_2$) is reported here. All the samples possess single-phase character and their superconducting transition temperatures ($T_c^{R=0}$) are 85 K, 90 K and 72 K respectively for Bi2212-MQ, Bi2212-$N_2$ and Bi2212-$O_2$. While Bi2212-MQ and Bi2212-$N_2$ samples are in near optimum doping regime, Bi2212-$O_2$ is an over-doped sample. $T_c^{S=0}$ values obtained through S(T) data are also in line with those deduced from the temperature dependence of resistance and DC magnetization. Interestingly, S(T) behaviour of the optimally-doped Bi2212-MQ and Bi2212-$N_2$ samples is seen to be positive in whole temperature range, it is found negative for the over-doped Bi2212-$O_2$ sample above $T_c^{S=0}$. These results have been seen in the light of the recent band structure calculations and the ensuing split Fermi surface as determined by angle-resolved photoelectron spectroscopy (ARPES).

---

* Corresponding Author: Dr. V.P.S. Awana , awana@mail.nplindia.ernet.in
www.freewebs.com/vpsawana/
Fax : 0091-11-45609310: Phone : 0091-11-45609210

## INTRODUCTION

Ever since the discovery of high $T_c$ superconductivity (HTSc) in perovskite cuprates [1], the theoretical understanding of the phenomenon is still elusive. The observance of the high superconducting transition temperatures of up to 132 K [2] and the anomalous normal state physical properties of the HTSc are quite intriguing [3-5]. Principally not only the pairing mechanism for very high $T_c$ has been undecided, precise normal state electrical or thermal conduction processes are also far from being clear [3-7]. Thermoelectric power S(T) behavior in these regards is one of the important tools. Unlike the resistivity R(T), the S(T) precisely mimics the exact electronic conduction process in terms of the dominating charge carriers [8,9]. For example in $NaCoO_2$ the electrical conduction is through highly metallic Co-O layers and the S(T) is controlled from phonon drag Na-O, giving rise to high conductivity along with the higher S value [10]. In case of HTSc cuprates, several articles exist in the literature on their S(T) but apparently no consensus seems to has been arrived at [4,11-14]. Though the R(T) behavior mostly match in various reports the S(T) does not [11-15]. This is precisely because of the split Fermi surface and ensuing two-band structure in most of HTSc [16-18]. Further, the distribution of oxygen in highly inhomogeneous HTSc samples, in particular the Bi-based cuprates, does hamper the real out put [19]. In case of Bi2212 though negative S(T) is reported in some reports for the over-doped samples [13], the same is also reported positive in others [20]. This could primarily be attributed to inhomogeneous oxygen distribution within the grains of polycrystalline samples [19].

In present article we address this problem, and present the S(T) results on oxygen homogenized Bi2212 samples from optimum- to over-doping regime. This is achieved by relatively longer hours annealing the samples in various environments. In particular, for obtaining the over-doped regime we annealed Bi2212 in $O_2$ atmosphere for around 36 hours. Also the S(T) is measured very precisely [21] and the repeatability of the experiments is checked. We found that the S(T) of over doped Bi2212 is unambiguously negative throughout its normal state i.e. above its $T_c$. This is also in contrast to one of our previous results on an in homogenously oxygen distributed sample [20]. The negative S(T) of the over-doped Bi2212 is in direct confirmation with the recent band structure calculations based on angle-resolved photoelectron spectroscopy (ARPES) studies [13,18].

## EXPERIMENTAL DETAILS

Samples of $Bi_2Sr_2CaCu_2O_8$ (Bi2212) were synthesized through the general solid-state reaction route using the ingredients $Bi_2O_3$, $CaCO_3$, $SrCO_3$ and CuO with more than 3N purity. Stoichiometric ratios were taken, ground thoroughly, and subsequently heat-treated thrice at 800°C for 24 hrs each in air with intermediate grindings. The bar shaped samples were then partially melted at 930°C for few minutes and annealed in the same furnace at 860°C for 20 hours before quenching to room temperature [22]. The samples thus produced are named as Bi2212-MQ. One of the samples from among Bi2212-MQ set is further annealed in flow of $N_2$ gas at 600°C for 20 hours and cooled slowly to room temperature in same atmosphere. This sample is named as Bi2212-$N_2$. Another sample from Bi2212-MQ set is annealed in $O_2$ gas flow at 600°C for 36 hours and cooled slowly to room temperature in same atmosphere. This sample is named as Bi2212-$O_2$. X-ray diffraction (XRD) patterns of the samples were recorded by using $CuK_\alpha$ radiation (Rigaku miniFlex-II, $\lambda$ = 1.54Å). R(T) measurements in the temperature range of 20 to 300K were carried out on a close cycle refrigerator using four-probe method. Temperature dependence of DC magnetization of the Bi2212-MQ and Bi2212-$O_2$ samples was carried out in a SQUID magnetometer (Quantum Design) both in zero field and field cooled configurations. The thermoelectric power S(T) measurements were carried out on an automated precision measurement system [21].

## RESULTS AND DISCUSSION

Figure 1(a) depicts the room temperature x-ray diffraction (XRD) patterns of Bi2212-MQ, Bi2212-$N_2$ and Bi2212-$O_2$ samples. All the samples exhibit phase purity with tetragonal structure. The characteristic low angle {002} peak is seen distinctly at $2\theta$ ~$5.8^0$. Any intermixing of Bi2212 phase (002 at $2\theta$ ~ $5.8^0$) with that of Bi2201 (002 at $2\theta$ ~$7.6^0$) or Bi2223 (002 at $2\theta$ ~ $4.6^0$) is also ruled out. The lattice parameters a and c of Bi2212-MQ and Bi2212-$O_2$ samples calculated with Rietveld refinement are 3.8238±

0.0004 Å, 30.8744± 0.0037 Å and 3.8177 ±0.0048 Å, 30.8457± 0.0038 Å respectively. The observed decrease in lattice parameters for Bi2212-$O_2$ sample vis-à-vis the Bi2212-MQ sample is clear indication of the increased oxygen content for the same sample. The decrease in c-parameter for Bi2212-$O_2$ sample is clear from its characteristic low angle (002) peak shift towards higher angle (see inset Fig.1a).

The resistance versus temperature plots R(T) for Bi2212-MQ, Bi2212-$N_2$ and Bi2212-$O_2$ samples are provided in Fig.1(b). The normal state conductivity of Bi2212-$O_2$ sample has improved tremendously in comparison to both Bi2212-MQ and Bi2212-$N_2$ samples, indicating a higher density of mobile carriers in Bi2212-$O_2$ material vis-à-vis Bi2212-MQ and Bi2212-$N_2$ samples. Interestingly, the superconductivity transition temperature ($T_c$) is least around 72K for the Bi2212-$O_2$ in comparison to 85K and 90K respectively for Bi2212-MQ and Bi2212-$O_2$ samples. This shows that Bi2212-$O_2$ annealed sample is over-doped, i.e., better normal state conductivity with lower $T_c$ in comparison to Bi2212-MQ and Bi2212-$N_2$ samples. DC magnetization measurements of the MQ and $O_2$ treated samples in both ZFC and FC configurations, also substantiate these $T_c$ values (Fig.2)

The thermoelectric power S(T) behavior of the studied Bi2212-MQ, Bi2212-$N_2$ and Bi2212-$O_2$ samples is given in main panel in Fig.3. All the three samples exhibit superconductivity with $T_c^{S=0}$ at nearly the same $T_c^{R=0}$ temperatures. As far as the observation of superconductivity is concerned both the R(T) and S(T) measurements corroborate each other (Fig.1(b) and Fig.3). However, there is one striking difference i.e., the normal state S(T) is though positive for Bi2212-MQ, Bi2212-$N_2$, the same is negative for Bi2212-$O_2$ sample. This behavior is anomalous in a single band scenario. However, the recent ARPES results do speculate such an outcome for the S(T) of over-doped Bi based cuprates in particular [16-18] and others in general [16]. On the other hand in case of over-doped YBCO ($YBa_2Cu_3O_7$) the observation of positive S(T) is discussed in terms of two band (Cu-O chain and plane) contributions. Very recently Kaminski et al [18] reported detailed ARPES studies and concluded that for heavily over-doped Bi2212 system the anti-bonding sheets become electron-like. It is shown that for number of holes per Cu-O plane in excess of 0.22, a crossover from hole to electron conduction takes place. In case of $(Bi,Pb)_2(Sr,La)_2CuO_6$, simultaneous ARPES and S(T) were performed by Kondo et al[13] and the thermo-power data was fitted from positive to negative spectrum for under- to over-doped samples. The recent ARPES studies [13,18] have clearly predicted the negative S(T) behavior for the heavily over-doped samples as has been observed in the present studies for the longer hours $O_2$ annealed over-doped Bi2212 sample. The primary cause for such a behavior was predicted quite some time back by McIntosh and Kaiser [16] in terms of a van Hove singularity in the electronic density of states of HTSc cuprate superconductors.

## CONCLUSION

In this short note, we report R(T), M(T) and S(T) data obtained on single phase Bi2212 samples synthesized via solid state reaction route with their $T_c$ values ranging from 90 to 72 K. Long hours of $O_2$ annealing facilitated in achieving the homogeneous over-doped regime of the compound with $T_c$ of 72 K. The observed negative S(T) dependence of the over-doped Bi2212 seems to be in accord with the recent band structure calculations and is in line with the theoretical investigations on a van Hove singularity in electronic density of states in HTSc compounds.

## ACKNOWLEDGEMENT

Authors thank Prof. Vikram Kumar, Director, National Physical Laboratory, New Delhi for his constant encouragement and support. They also thank their colleague Dr. S.K.Agarwal for fruitful discussions. One of us (JKB) is further thankful to CSIR for financial help in terms of NET-JRF fellowship.

**FIGURE CAPTIONS**

Fig.1(a) X-ray diffractograms of the various Bi2212 samples, the inset shows the extended low angle characteristic (002) peak.

Fig.1(b) Temperature dependence of resistance R(T) of the Bi2212-MQ, Bi2212-$N_2$ and Bi2212-$O_2$ samples.

Fig.2 Temperature dependence of DC magnetization of the Bi2212-MQ and Bi2212-$O_2$ samples in ZFC and FC configurations.

Fig.3 Temperature dependence of thermopower S(T) of the Bi2212-MQ, Bi2212-$N_2$ and Bi2212-$O_2$ samples.

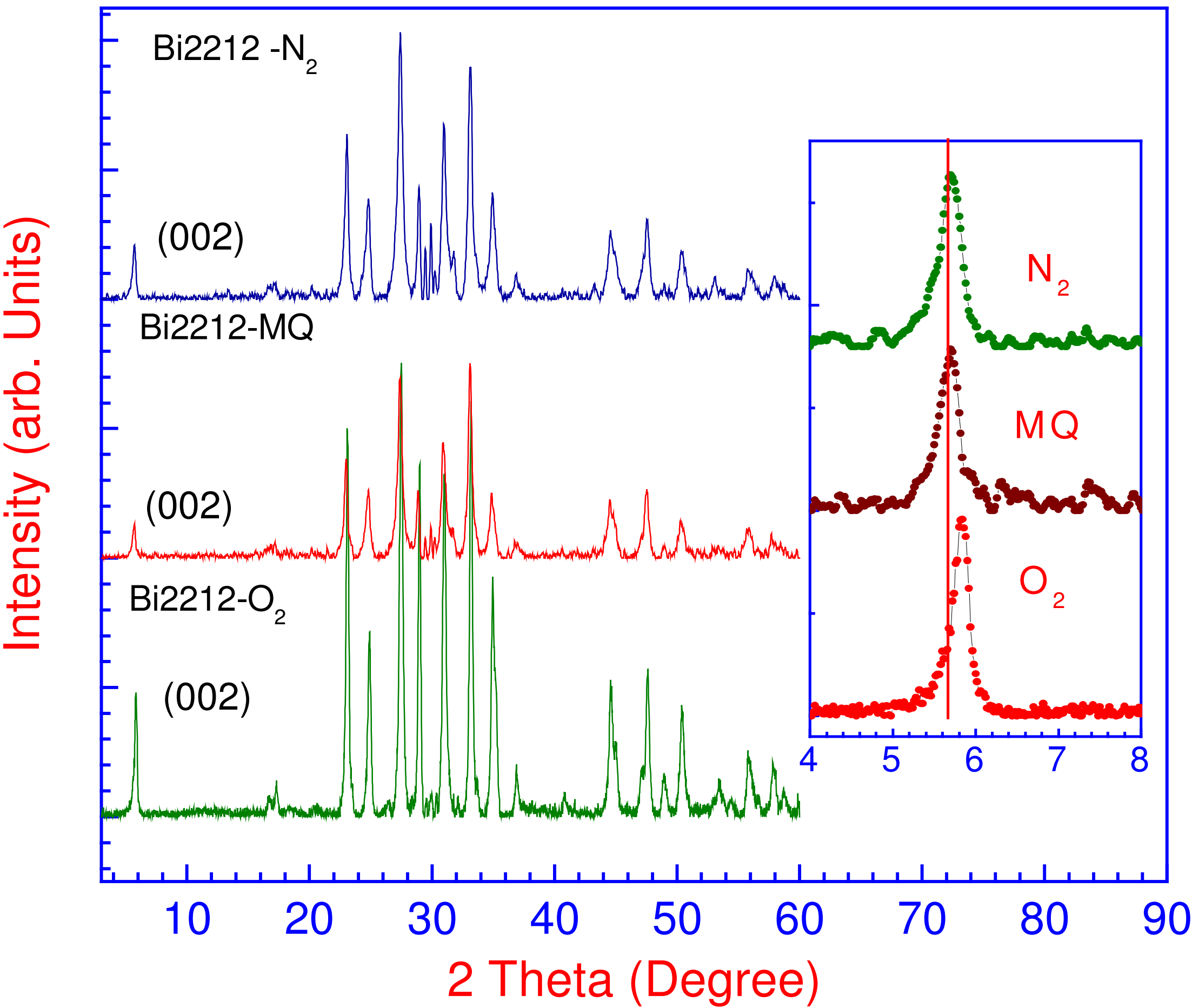

**Figure 1(a)**

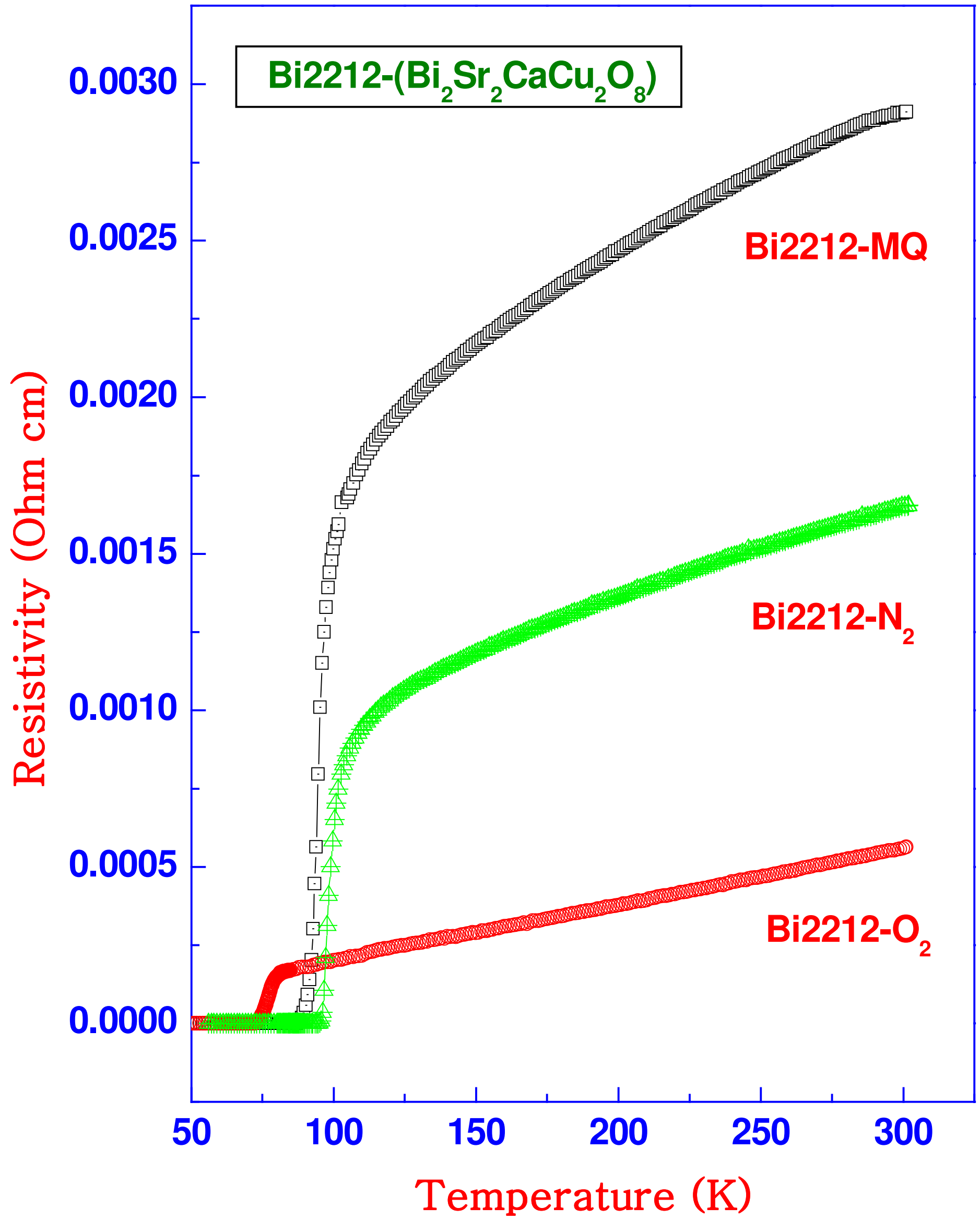

**Figure 1(b)**

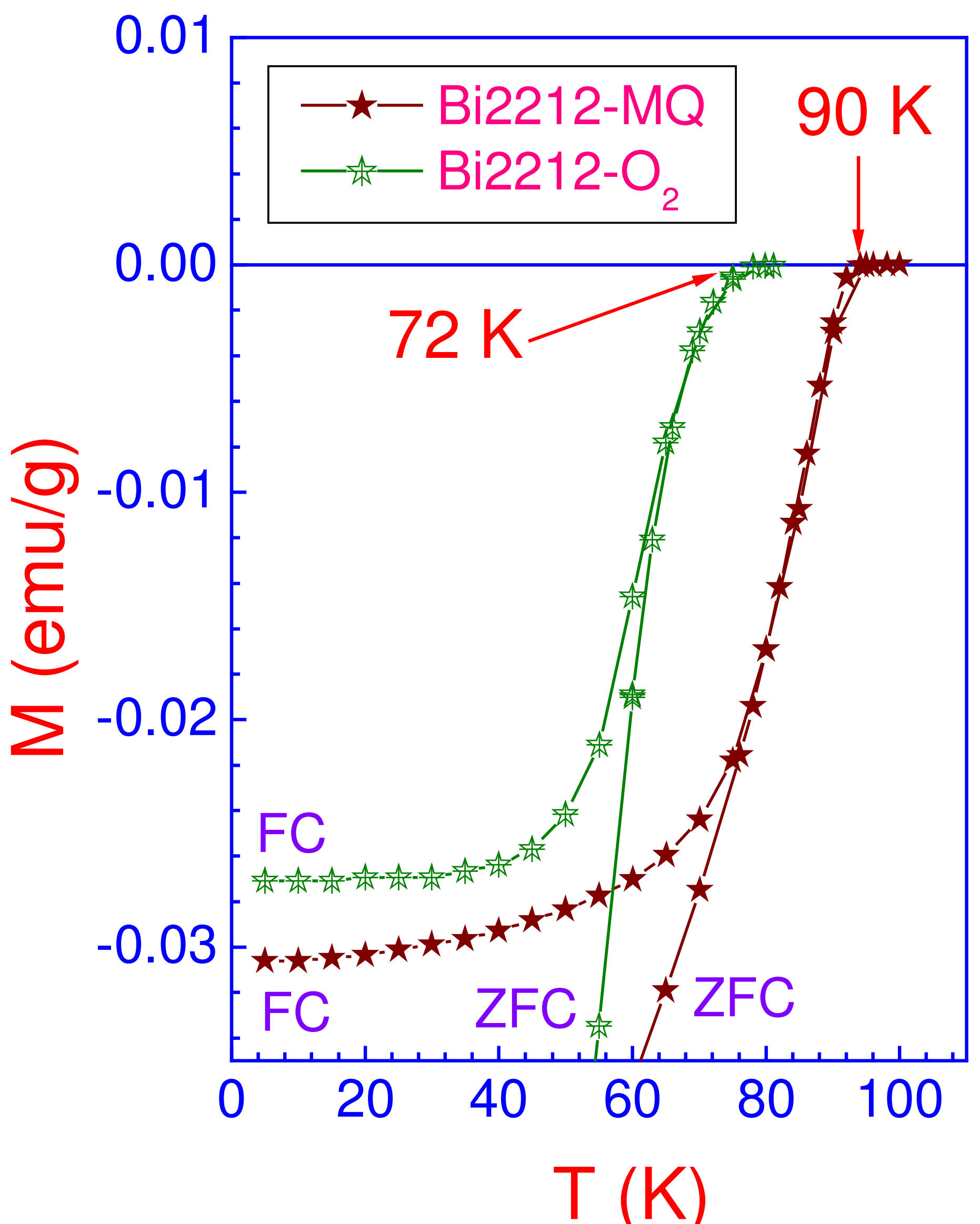

**Figure 2**

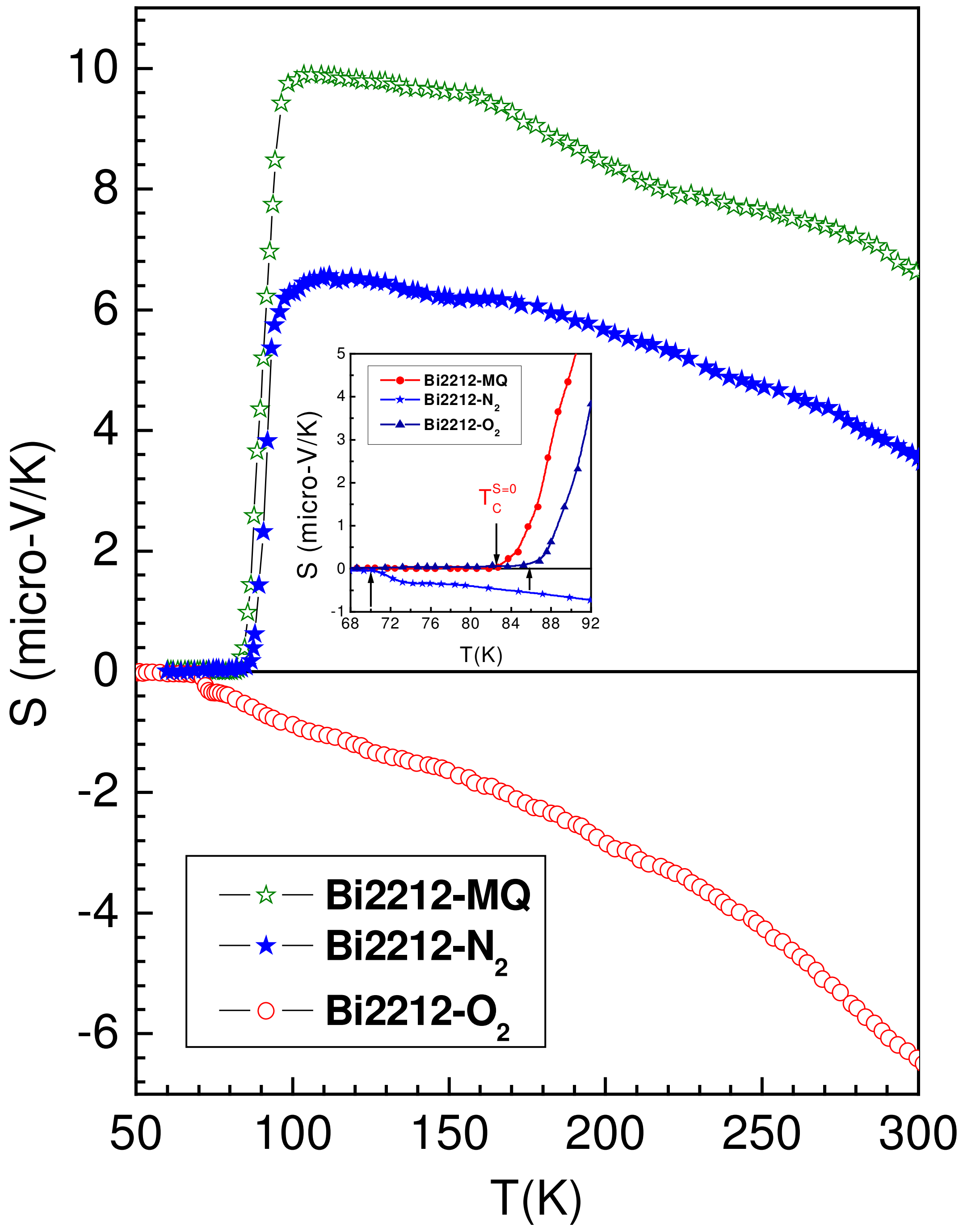

**Figure 3**